\documentclass[11pt]{article}
\usepackage{moriond,epsfig,amssymb,subfigure}

\newcommand{\omqls}{\Omega_Q^{ls}}
\newcommand{\mmc}{MCMC}

\def\be{\begin{equation}}
\def\ee{\end{equation}}
\def\bea{\begin{eqnarray}}
\def\eea{\end{eqnarray}}

\begin{document}
\vspace*{4cm}
\title{Recent Constraints on Models of Quintessence}

\author{M.Doran}

\address{Department of Physics \& Astronomy, Dartmouth College,
6127 Wilder Laboratory,
Hanover, NH 03755}

\maketitle\abstracts{
We perform a comparison of the WMAP measurements  
with the predictions of quintessence cosmological
models of dark energy. We consider a wide range of quintessence models,
including: a constant equation-of-state;  a simply-parametrized, time-evolving
equation-of-state;  a class of models of early quintessence; scalar fields with
an inverse-power law potential. We also provide a joint fit to the CBI and
ACBAR CMB data, and the type 1a supernovae.}

\begin{figure}[t]
\begin{center}
\includegraphics[scale=0.45]{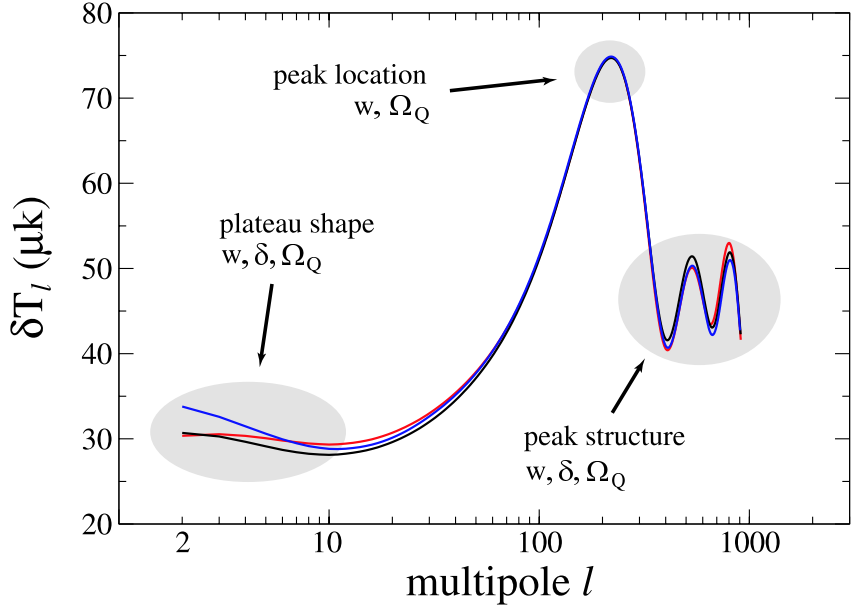}
\caption{\label{fig:cmb_bw} 
The pattern of CMB anisotropy can reveal information about the quintessence
abundance ($\Omega_Q$), equation-of-state ($w$), and behavior of fluctuations
($\delta$). The three curves are examples of constant equation-of-state models
which differ little by eye, but are distinguished by the data. The red
($w=-0.5$) and blue ($-1.2$) curves are both low-$\chi^2$
CMB-indistinguishable, but distinct with respect to SNe. The black curve
($-0.8$), although it is consistent with the SNe data 
is rejected by the CMB at the $3\sigma$-level.} 
\end{center}
\end{figure}

\section{Introduction}
We carry out an extensive analysis \cite{Caldwell:2003hz} of the CMB anisotropy and
mass fluctuation spectra for a wide range of quintessence\cite{Ratra:1987rm,Wetterich:fm,Caldwell:1997ii}
models. These models
are: (Q1) models with a constant equation-of-state, $w$, including $w<-1$; (Q2)
models with a simply-parametrized, time-evolving $w$; (Q3) early quintessence
models, with a non-negligible energy density during the recombination era; (Q4)
trackers described by a scalar field evolving under an inverse-power law
potential.  


The suite of parameters describing the cosmological models are split into
quintessence parameters, $\theta_{Q}$ and 
spacetime plus ``matter sector" variables, $\theta_{M}$, where $\theta_{M} =
\{\Omega_b h^2,$ $\Omega_{cdm} h^2,\, h,\, n_s,$ $A_S,\,\tau_{r}\}$. In order,
these are the baryon density, cold dark matter density, hubble parameter,
scalar perturbation spectral index, scalar perturbation amplitude, and optical
depth. We restrict our attention to spatially-flat, cold
dark matter models with a primordial spectrum of nearly scale-invariant density
perturbations generated by inflation.  \\[-3ex]
\section{Constant $w$ and phantom dark energy}
For a constant equation of state,
we have used the equivalence between a scalar field $\varphi$ with
potential $V(\varphi)$ and the equation-of-state $w$ in order to
self-consistently evaluate the quintessence fluctuations. For the range $w<-1$
we employ a {\it k}~essence model, keeping the sound speed (actually, this is
$d\omega^2/dk^2$) fixed at $c_s^2 = 1$. Since this model introduces only one
additional parameter beyond the basic set of spacetime plus matter sector
variables, we adopt a simplistic grid-based search for viable models. The
acceptance criteria for the Q1 models is based on a $\Delta\chi^2$-test. The
results of our survey of Q1 models are shown in
Figures~\ref{fig:HubbleFit},\,\ref{fig:BestFit}. Our basic conclusion from the overlapping
constraint regions is that there exist concordant models with $-1.25 \lesssim w
\lesssim -0.8$ and $0.25 \lesssim \Omega_{m} \lesssim 0.4$.

\newcommand{\twowidth}{7.7cm}
\newcommand{\twoheight}{5cm}
\begin{figure}[t]
\subfigure[The one-parameter family of best-fit models, which exploit the geometric
degeneracy of the CMB anisotropy pattern, is shown as the thick, red curve in
the $w-h$ plane. We have explored models in a six-dimensional cylinder in the
parameter space surrounding this ``best-fit line.''  The HST-Key Project
$1\sigma$ measurement of the Hubble constant is shown by the shaded band. ]{\label{fig:HubbleFit}\includegraphics[width=\twowidth,height=\twoheight]{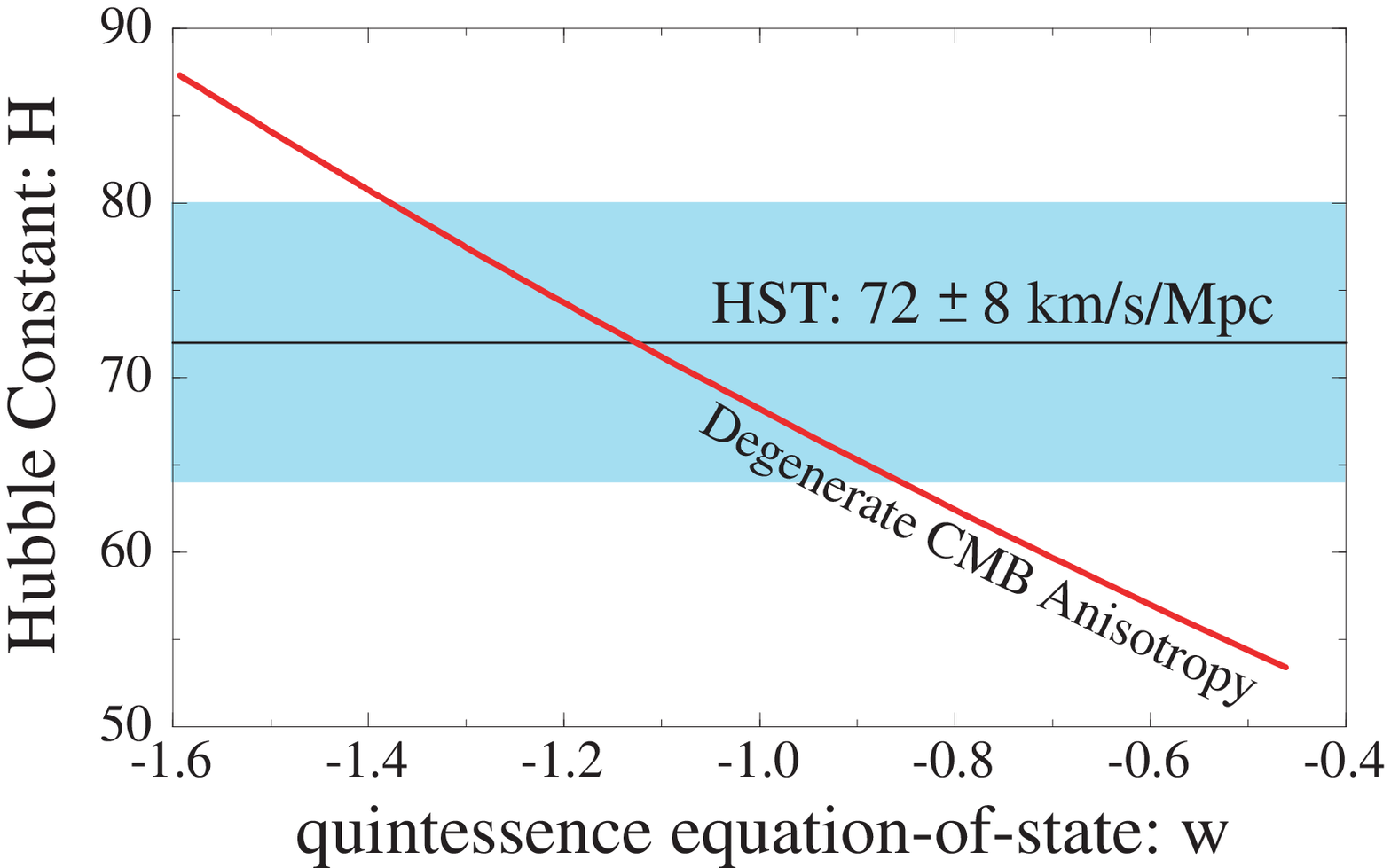}}
\subfigure[The constraints on constant equation-of-state models due to CMB (WMAP, ACBAR,
CBI) and type 1a supernovae (Hi-Z, SCP) are shown. The starting point for our
parameter-search, the family of CMB-degenerate models, is shown by the thick,
black line.]{\label{fig:BestFit}\includegraphics[width=\twowidth,height=\twoheight]{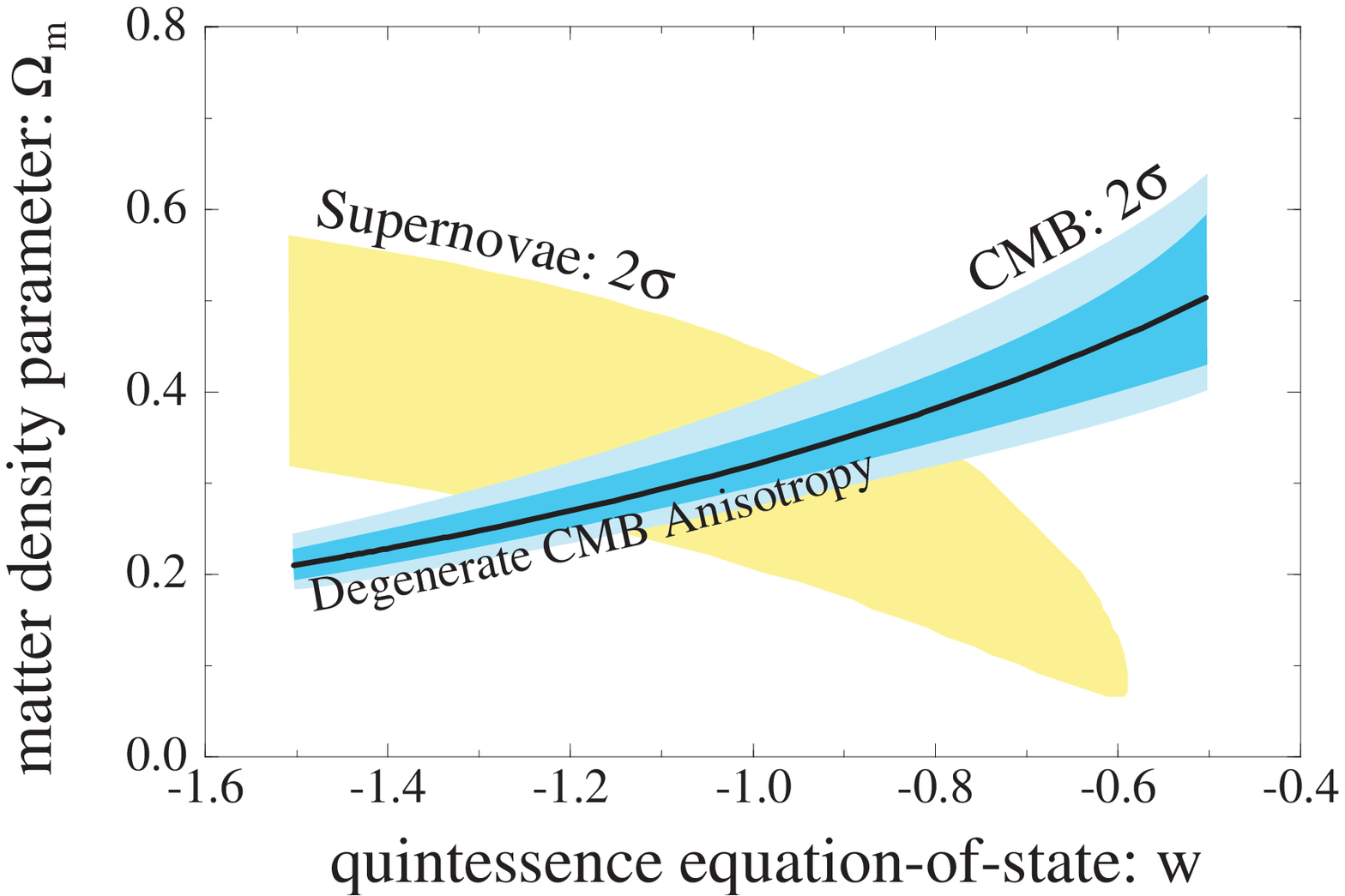}}
\end{figure}


\newcommand{\threescale}{0.22}
\begin{figure*}[t]
\subfigure[Q2: monotonic evolution $w(a)$]{\label{fig::monotonic1}
  \includegraphics[scale =\threescale,angle=-90]{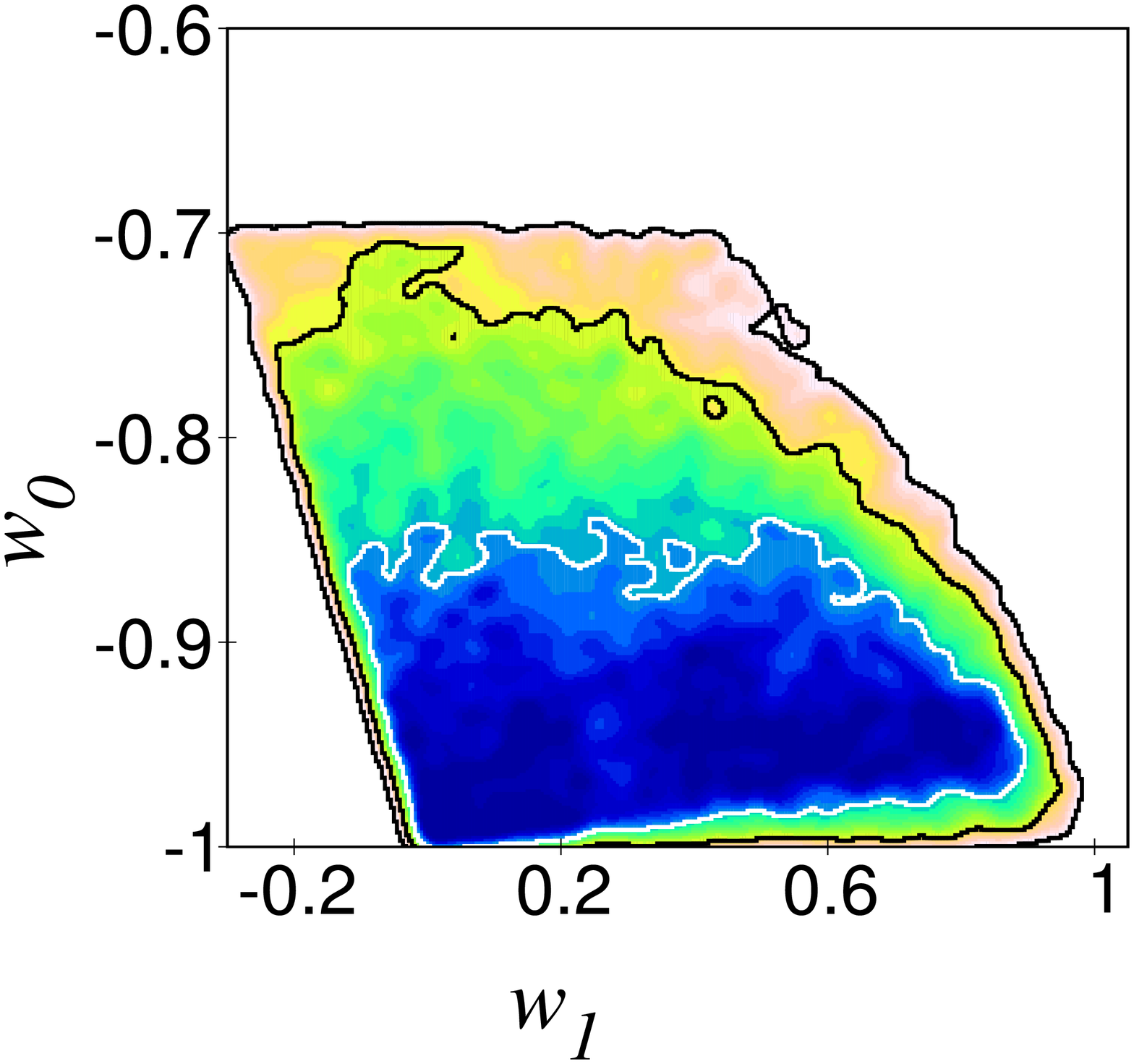}
}
\subfigure[Q3: leaping kinetic quintessence]{ \label{fig::leaping1}
  \includegraphics[scale =\threescale,angle=-90]{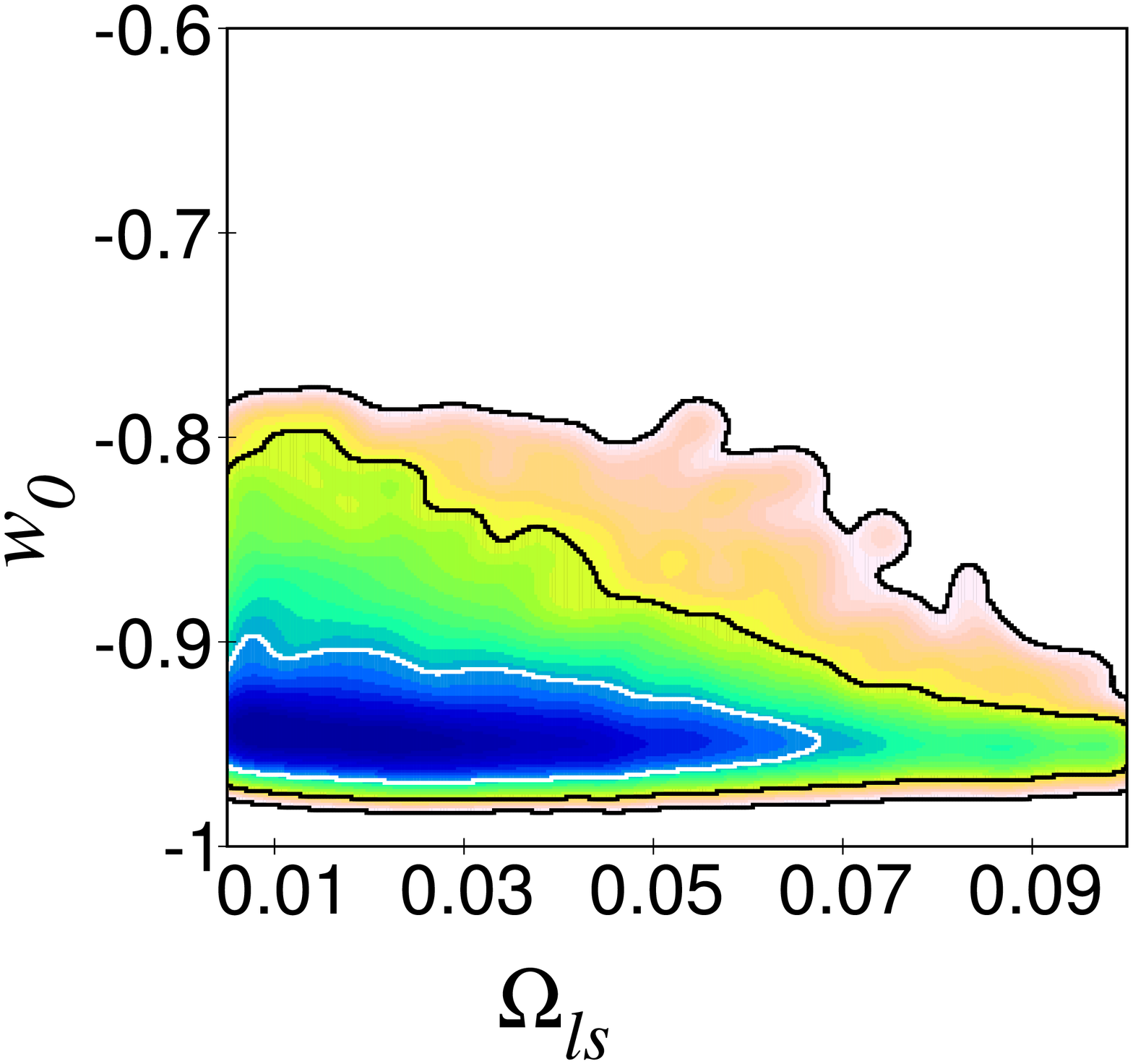}
}
\subfigure[Q4: inverse-power law]{ \label{fig::ipl1}
  \includegraphics[scale =\threescale,angle=-90]{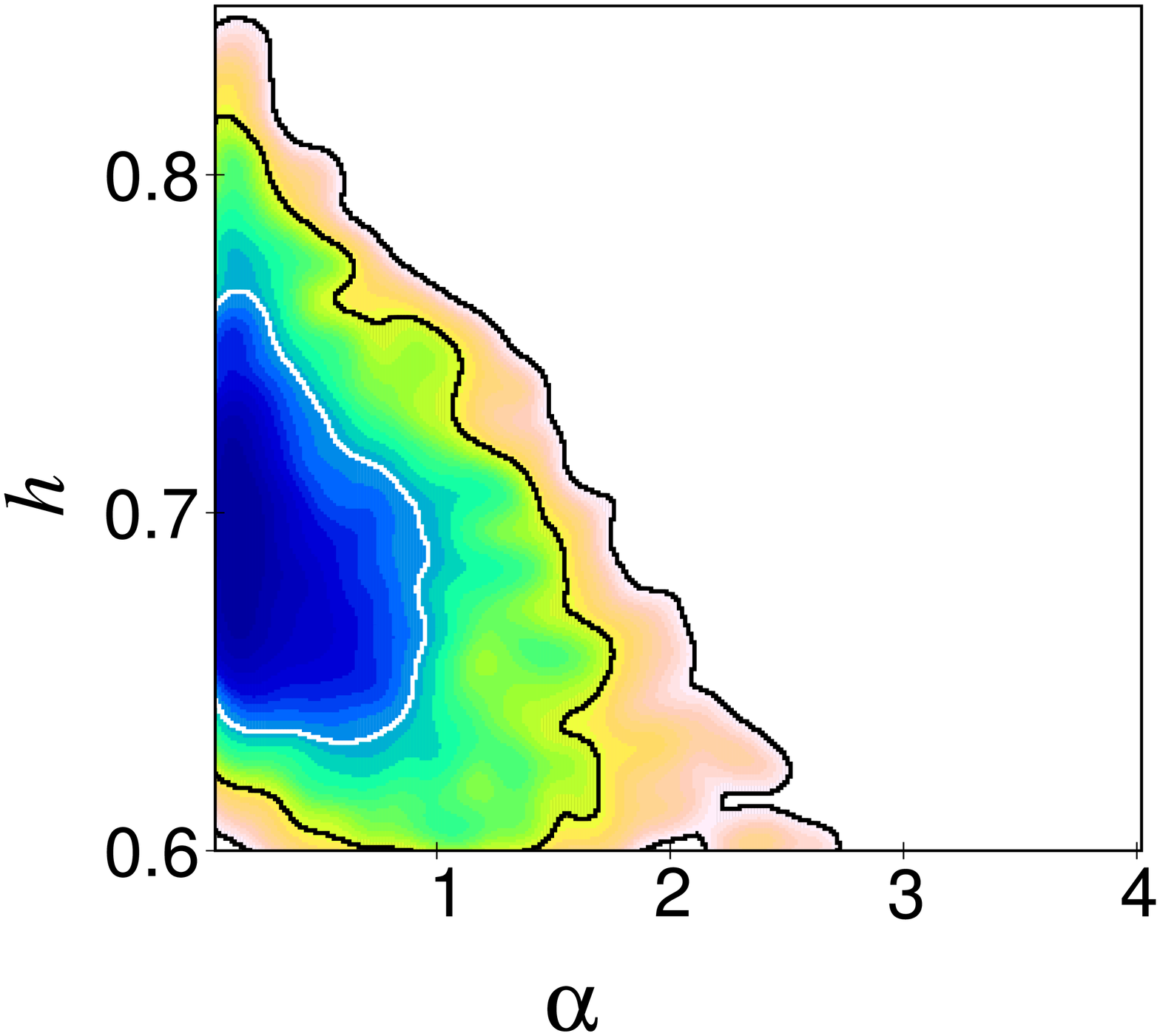}
}

\subfigure[Q2: monotonic evolution $w(a)$]{ \label{fig::monotonic2}
  \includegraphics[scale =\threescale,angle=-90]{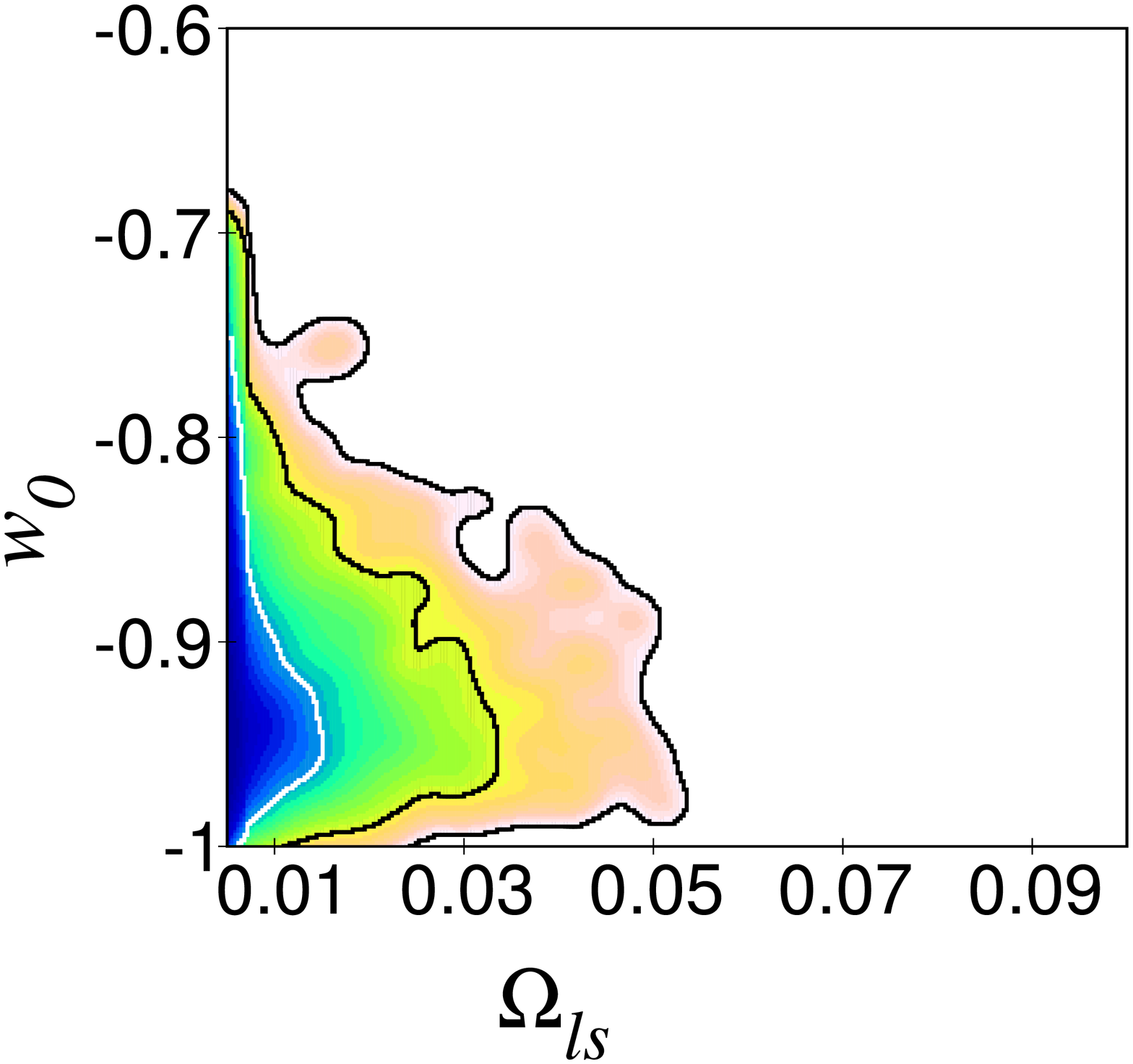}
}
\subfigure[Q3: leaping kinetic quintessence]{ \label{fig::leaping2}
  \includegraphics[scale =\threescale,angle=-90]{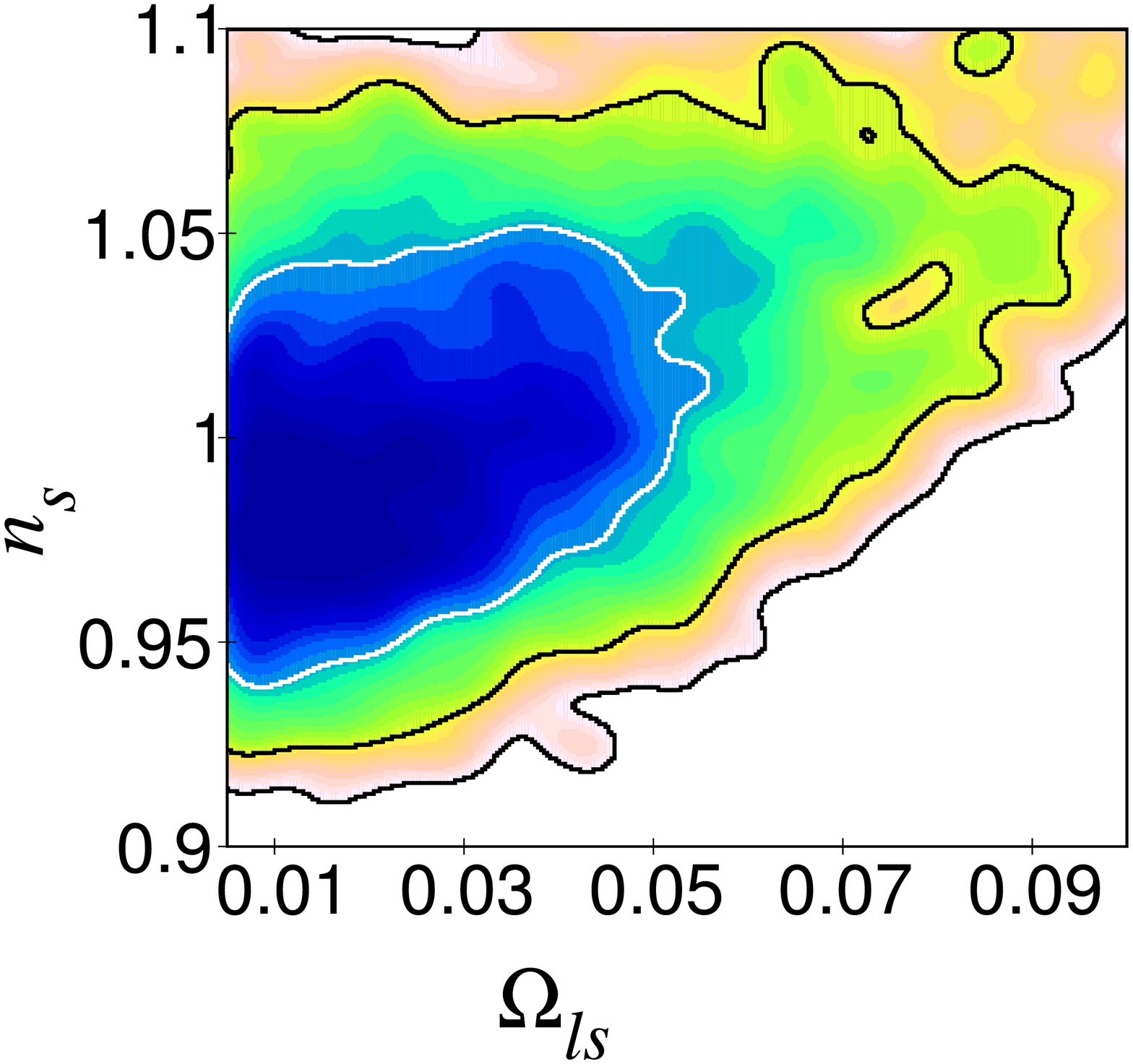}
}
\subfigure[Q4: inverse-power law]{\label{fig::ipl2}
  \includegraphics[scale =\threescale,angle=-90]{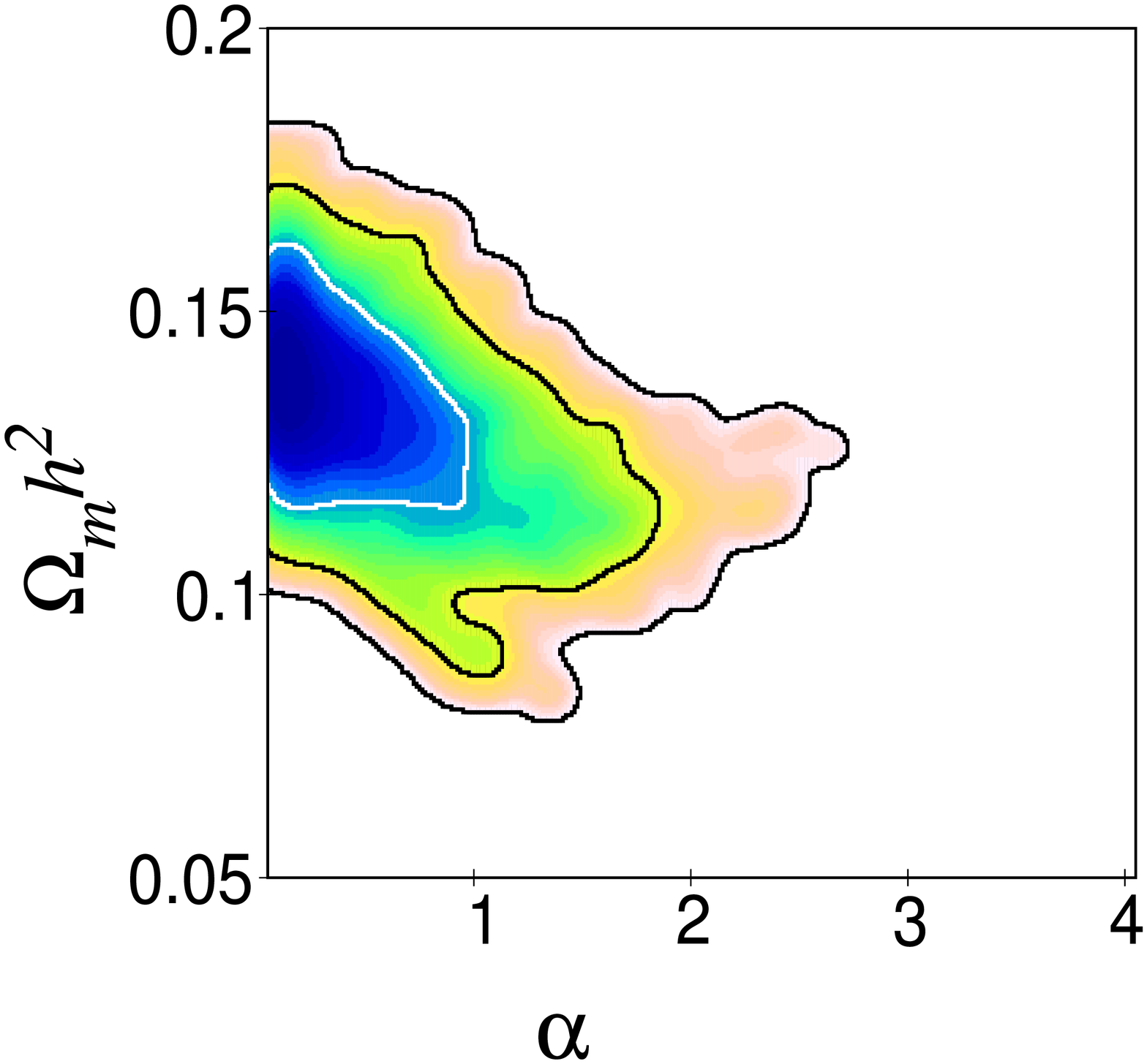}
}
\caption{The results of our \mmc\ search of the multi-dimensional parameter
search, for models Q2-4, are illustrated in the figures above. In all cases, we
have marginalized over the suppressed parameters. The solid lines indicate the
$1,\,2,\,3\sigma$ contours based on comparison with the CMB (WMAP, ACBAR, CBI)
and type 1a supernovae (Hi-Z, SCP).} 
\end{figure*}

\section{Celestine quintessence:  $w(a) =w_0 + (1-a) w_1$}
We have examined quintessence models with an
equation-of-state that evolves monotonically with the scale factor, as $w(a) =
w_0 + (1-a) w_1$. This parametrization has been shown to be versatile in
describing the late-time quintessence evolution for a wide class of scalar
field models\cite{Linder:2002et}. Based on the degeneracy of models found for
Q1, we expect to find a two-dimensional family of equivalent best-fit models
with the same apparent angular size of the last scattering horizon, occupying a
plane in the $\{w_0,\,w_1,\,h\}$ space.  We find $w_0 < -0.75$ at the $2\sigma$
level, marginalizing over the suppressed five dimensional parameter space, as
illustrated in Figure~\ref{fig::monotonic1}. There, the shapes of the contours
indicate that current data can only distinguish between fast ($dw/da \gtrsim
0.5$) and slow evolution of $w(a)$, and offer only a weak bound on $w_1$.
However, in terms of $\omqls$ the relative quintessence density during recombination, 
we find $\omqls < 0.03$ at the $2\sigma$ level.


\section{Leaping kinetic term quintessence}
Leaping kinetic
quintessence features a 
non-canonical kinetic term that undergoes a sharp transition at late times,
leading to the current accelerated expansion\cite{Hebecker:2000zb}. At early
times the field closely tracks the cosmological background with $w=0$ during
matter domination, appearing as early quintessence\cite{Caldwell:2003vp}
before undergoing a steep transition towards a strongly negative
equation-of-state by the present day.  Since
$\omqls$ is not tied as closely to the expansion rate sampled by the
supernovae, compared to case Q2, the result is the weaker constraint $\omqls
\lesssim 0.1$, as shown in Figure~\ref{fig::leaping1}. Although the limit of a
cosmological constant can be approached in this model, the presence of early
quintessence will then require a sharp transition in the equation-of-state in
order to reach $w \to -1$. Consequently, we bar
models with $w \approx -1$ and non-negligible $\omqls$, as displayed in
Figure~\ref{fig::leaping1}. Next, because early quintessence
suppresses  the growth of fluctuations on small scales compared to large
scales, we find that comparable fluctuation spectra can be achieved by making a
trade-off between $n_s$ and $\omqls$ (see Figure~\ref{fig::leaping2}).\\[-4.5ex]


\section{Inverse power law quintessence}
Inverse-power law (IPL) models are the archetype quintessence
models with tracking property and acceleration\cite{Ratra:1987rm,Zlatev:1998tr,Steinhardt:nw}.  The potential is given by $V \propto
\varphi^{-\alpha}$, where the constant of proportionality is determined by
$\Omega_Q$. In certain supersymmetric QCD realizations of the IPL
\cite{Masiero:1999sq}, $\alpha$ is related to the numbers of color and flavors,
and can take on a continuous range of values $\alpha >0$. For $\alpha \to 0$,
however, inverse-power law models behave more and more like a cosmological
constant. From our analysis, we see that $\alpha \lesssim 1 - 2$, only a minor improvement
of earlier investigations\cite{Balbi:2000kj,Baccigalupi:2001aa,Doran:2001ty} using pre-WMAP data.

In   Figure~\ref{fig::ipl2} we plot the
likelihood contours in the $\Omega_m h^2 - \alpha$ plane: our results agree
with the best fit at $\Omega_m h^2 = 0.149$ for $\alpha=0$ or $w=-1$, but show
a tolerance for a wider range for $0 \le \alpha \le 2$. That is, the additional
degree of freedom in $\alpha$ means that the matter density for the IPL model
is not as well-determined from the peak position \cite{Page:2003fa} as compared
to the $\Lambda$ model. However, to maintain the peak at $\ell = 220$, we
observe that $\Omega_m h^2$ decreases slightly as $\alpha$ increases. \\[-3ex]

\section*{Acknowledgments}
This work was supported by NSF grant
PHY-0099543 at Dartmouth. We thank Pier Stefano Corasaniti for useful
conversations, and Dartmouth colleagues Barrett Rogers and
Brian Chaboyer for use of computing resources.


\begin{thebibliography}{99}


\bibitem{Caldwell:2003hz}
R.~R.~Caldwell and M.~Doran,
arXiv:astro-ph/0305334.

 
\bibitem{Ratra:1987rm}
B.~Ratra and P.~J.~Peebles,
Phys.\ Rev.\ D {\bf 37}, 3406 (1988).

 
\bibitem{Wetterich:fm}
C.~Wetterich,
Nucl.\ Phys.\ B {\bf 302}, 668 (1988).



\bibitem{Caldwell:1997ii}
R.~R.~Caldwell, R.~Dave and P.~J.~Steinhardt,
Phys.\ Rev.\ Lett.\  {\bf 80}, 1582 (1998).


\bibitem{Linder:2002et}
E.~V.~Linder,
Phys.\ Rev.\ Lett.\  {\bf 90}, 091301 (2003).

\bibitem{Hebecker:2000zb}
A.~Hebecker and C.~Wetterich,
Phys.\ Lett.\ B {\bf{497}}, 281 (2001). 

\bibitem{Caldwell:2003vp}
R.~R.~Caldwell, M.~Doran, C.~M.~Mueller, G.~Schaefer and C.~Wetterich,
[arXiv:astro-ph/0302505].

\bibitem{Zlatev:1998tr}
I.~Zlatev, L.~M.~Wang and P.~J.~Steinhardt,
Phys.\ Rev.\ Lett.\  {\bf 82}, 896 (1999).

\bibitem{Steinhardt:nw}
P.~J.~Steinhardt, L.~M.~Wang and I.~Zlatev,
Phys.\ Rev.\ D {\bf 59}, 123504 (1999).

\bibitem{Masiero:1999sq}
A.~Masiero, M.~Pietroni and F.~Rosati,
Phys.\ Rev.\ D {\bf 61}, 023504 (2000).

\bibitem{Balbi:2000kj}
A.~Balbi, C.~Baccigalupi, S.~Matarrese, F.~Perrotta and N.~Vittorio,
Astrophys.\ J.\  {\bf 547}, L89 (2001).

\bibitem{Baccigalupi:2001aa}
C.~Baccigalupi, A.~Balbi, S.~Matarrese, F.~Perrotta and N.~Vittorio,
Phys.\ Rev.\ D {\bf 65}, 063520 (2002).

\bibitem{Doran:2001ty}
M.~Doran, M.~Lilley and C.~Wetterich,
Phys.\ Lett.\ B {\bf 528}, 175 (2002).


\bibitem{Page:2003fa}
L.~Page {\it et al.},
arXiv:astro-ph/0302220.

\end{thebibliography}
\end{document}